\documentclass[12pt]{article}
\usepackage[T1]{fontenc}
\usepackage{amssymb,amsthm,amsmath,mathrsfs,bm}

\begin{document}

\title{ Double Extended  Cubic Peakon  Equation}
\author{ Z. Popowicz}

\maketitle

\begin{center}{
Institute of Theoretical Physics, University of Wroc\l aw,

Wroc\l aw pl. M. Borna 9, 50-205 Wroc\l aw Poland, ziemek@ift.uni.wroc.pl}
\end{center}
\vspace{0.8cm} 

\begin{abstract}
The Hamiltonian structure for the supersymmetric $N=2$ Novikov equation is presented. The bosonic sector give us two-component generalization of the cubic peakon equation. The double extended: two component and two peakon Novikov equations are defined.  
The Bi-Hamiltonian structure for this extended system is constructed. 
\end{abstract}
\bigskip

\section{Introduction}

Recently a family of equations of the form [1 - 7] 
\begin{equation}
u_t - u_{xxt}  = \frac{1}{2} \big ( -(b+1) u^2 + 2uu_{xx} + (b-1) u_x^2 \big )_x
\end{equation}
has been investigated in the literature . 

When $ b=2$ Eq.(1) reduces to the Camassa - Holm equation
\begin{equation}
u_t - u_{xxt}  = \frac{1}{2} \big ( -3 u^2 + 2uu_{xx} +  u_x^2 \big )_x
\end{equation}
 which  describes a special approximation of shallow water theory. This equation shares most of the 
important properties of an integrable system of KdV type, for example, the existence of Lax pair formalism, the 
Bi-Hamiltonian structure, the multi-solitons solutions. Moreover this equation admits peaked solitary wave solutions.

Degasperis and Procesi showed that the Eq. (1)  is integrable also for the $b=3$ case. 
The Degasperis - Procesi equation 
\begin{equation}
u_t - u_{xxt} = (-2u^2 + uu_{xx} + u_x^2 )_x 
\end{equation}
can be considered as a model for shallow-water dynamics also  and found to be completely integrable. 
Similarly to the Camassa-Holm case the Degasperis-Procesi equation has the Lax pair and admits 
peakon dynamics also. 

The peakon equation  have been generalized  to the so called cubic peakon equation by Novikov \cite{Novikov} and one of them  is
\begin{equation}\label{novikov}
m_{t}+u^{2}m_{x}+3uu_{x}m=0,\quad m=u-u_{xx}.
\end{equation}
Hone and Wang \cite{Hone} proposed a Lax representation for the equation (\ref{novikov}) and found the Bi-Hamiltonian structure and  infinitely many conserved quantities .

The Camassa-Holm, Degasperis-Procesi and Novikov  equations have been generalized to the multi-component case in different manners. For example 
the energy - dependent Schr{\"o}dinger spectral problem for Camassa-Holm equation can be formulated with the help of Lax operator as
\begin{eqnarray}\label{laxch}
\Psi_{xx} &=& (\frac{1}{4}  - \lambda m) \Psi \\ \nonumber 
\Psi_t &=& - (\frac{1}{2\lambda} + u )\Psi_{x} + \frac{1}{2}u_x\Psi
\end{eqnarray}
 Compatibility condition for the above system, yields two independent equations 
\begin{equation}
m_t = -2mu_x -m_xu , ~~~~~~~ m = u - u_{xx}
\end{equation}

Considering now the class of the  two-component Schr{\"o}dinger equation {\cite{Rose,Zhang,Aratyn}} 
\begin{equation}
\Psi_{xx} = (\frac{1}{4}  - \lambda m + \lambda^2\rho^2) \Psi 
\end{equation}
the compatibility condition $\Psi_{x,t} = \Psi_{t,x}$,  where $\Psi_t$ is the same as in (\ref{laxch}),   yields equation 
\begin{eqnarray}\label{comp}
&& \rho_t=-(u\rho)_t , \\ \nonumber 
&& m_t = -2mu_x -m_xu + \rho\rho_x
\end{eqnarray}
This  equation  is known as the two-component generalization of the Camassa-Holm equation and is reducible to the Camassa-Holm equation when $\rho=0$.

The term \textit{multi-component generalization} of Camassa-Holm equation \cite{Zhang,QU} has been used also  to systems  which  are  reducible to scalar Camassa-Holm type equations.

A two-peakon generalization  of  the  Novikov equation was constructed by Geng and Xue \cite{Geng}.
\begin{eqnarray}\label{nu}
&& m_{t}+3u_{x}vm+uvm_{x} =0, \nonumber\\
&&n_{t}+3v_{x}un+uvn_{x} =0,  \\
&& m=u-u_{xx},\quad n=v-v_{xx}. \nonumber
\end{eqnarray}
which is reducible to the Eq. (\ref{novikov})  when $v=u,n=m$. 
They  calculated the $N$-peakons and conserved quantities and found a Hamiltonian structure.
The Bi-Hamiltonian structure have been found by Li and Liu \cite{Liu}.
Other types of the cubic peakon equation have been extended to the multi-peakon equation  \cite{Qiao1} also.

As we see we meet with   two different types of the generalization  of the pekon equation.  Therefore it is reasonable to introduce different terms on different types  of the extensions  of the these equations. 
If it is possible to reduce each component of the system of equations to the scalar peakon equation then such system of equation we name as multi-peakon generalization. 
For other types of the generalizations we reserve the term  multi-component extension.

The  Degasperis-Procesi equation have been generalized to the two-component \cite{Pop1} case using the supersymmetric technique but it appeared that it has one-Hamiltonian structure  only. It is possible to obtain the two-component Camassa-Holm equation \cite{Pop2} using the supersymmetric approach.

The aim of the present paper is twofold. The first is to construct the two-component Novikov equation, the second is to builds  its two-peakon extension. This type of equations we will name as double extended  cubic peakon equations. 

In order to define two-component Novikov equation we use, similarly to the Degasperis-Procesi equation, 
the extended $N=2$ supersymmetric and decompression technique described in \cite{Pop1}. 
As the result we obtained $N=2$ supersymmetric Novikov equation which bosonic sector give us two-component generalization of Novikov equation.

Next we  apply the decompression technique to the two-peakon Novikov equation and we obtained double extended  Novikov equation, two-peakon and two-component system.

The paper is organised as follows. The first section contains the supersymmetric approach in which we also explain the decompression technique. In the second section we constructed the doubly extended Novikov equation and presented its Bi-Hamiltonian structure. The last section contains concluding remarks. In the appendix we presented main tricks used in the computations of the Jacobi identity for the second Hamiltonian operator of the doubly extended Novikov equation.

\section{Supersymmetric Novikov Equation}

We will use  the  extended $N=2$ supersymmetric formalism which allows us to consider the
supersymmetric analog of the second Hamiltonian operator of the Novikov equation. 

Here we will use the supersymmetric algebra of super-derivatives where
\begin{eqnarray} 
&& {\cal D}_1 =\frac{\partial }{\partial \theta_1} -\frac{1}{2} \theta_2 \partial_x, ~~~~~~~~~
{\cal D}_2 =\frac{\partial }{\partial \theta_2} -\frac{1}{2} \theta_1 \partial_x, \\ \nonumber 
&& \{ {\cal D}_1 , {\cal D}_2 \}  = -\partial_x , ~~{\cal D}_3=  [{\cal D}_1, {\cal D}_2 ]= \partial_x + 2{\cal D}_1{\cal D}_2, ~~~ {\cal D}_1^2 = {\cal D}_2^2=0
\end{eqnarray}

Now let us consider the following Hamiltonian operator 
\begin{equation} \nonumber 
{\cal J}=\left ( \begin{array}{cc} -{\cal D}_3 \partial_x + 2\partial_x W + 2({\cal D}_1 W) {\cal D}_2 + 
2({\cal D}_2 W) {\cal D}_1 
 &  \partial_x V + 2({\cal D}_1 V) {\cal D}_2 + 2({\cal D}_2 V) {\cal D}_1 
 \\ \partial_x V + V_x + 2({\cal D}_1 V) {\cal D}_2 + 2({\cal D}_2 V) {\cal D}_1  & 0 \end{array} \right )
\end{equation}
where $W,V$ are the $N=2$ supersymmetric bosonic functions.

The super functions $W,V$  can be thought as the $ N=2$  supermultiplets,  for example $W(x, t, \theta_1, \theta_2) = w_o + \theta_1\chi_1 + \theta_2 \chi_2 + \theta_2\theta_1w_1 $
where $w_o(x,t), w_1(x,t)$  are the classical functions, $\chi_1(x,t), \chi_2(x,t)$ are Grassman valued
functions and $\theta_1, \theta_2$ are the Majorana spinors.

It is easy to check that this operator  satisfy the Jacobi identity. 

Let us briefly explain the standard Dirac reduction formula \cite{Dirac}. Let $U,V$ be two linear spaces with the coordinates $u$ and $v$. Let $P(u,v)$ be a Poisson tensor on $U \oplus V$ 
\begin{equation}
 P(u,v) = \left ( \begin{array}{cc} P_{u,u} & P_{u,v} \\ P_{v,u} & P_{v,v} \end{array} \right ).
\end{equation}
Assume that $P_{v,v}$ is invertible, then 
\begin{equation}
 P=P_{u,u} - P_{u,v} P_{v,v}^{-1} P_{v,u} 
\end{equation}
is a Poisson tensor on $U$ . 

Performing the Dirac reduction for ${\cal J}$ when   $W=1$ and taking into an account that 
\begin{equation}
 (-{\cal D}_3 + 2)^{-1} = (4-\partial_{xx})^{-1}(2+{\cal D}_3)
\end{equation}
we obtain  new Hamiltonian operator which satisfy the Jacobi identity.
\begin{eqnarray}\nonumber 
 && {\cal K} =  [ \partial_x V + V_x + 2({\cal D}_1 V) {\cal D}_2 + 2({\cal D}_2 V) {\cal D}_1 ]
(4\partial_x-\partial_{xxx})^{-1} (2+{\cal D}_3) \\ \nonumber 
&& \hspace{2cm} [\partial_x V + 2({\cal D}_1 V) {\cal D}_2 + 2({\cal D}_2 V) {\cal D}_1 ]
\end{eqnarray}

Let us now parametrise the superfunctin $V$ as $V=(1- {\cal D}_3)A$ where $A$ is a new supersymmetric function and consider the following equation
\begin{equation}
 V_t={\cal K}\frac{\delta H}{\delta V}
\end{equation}
Choosing $ H=\frac{1}{2} \int ~ dx d\theta_1 d\theta_2  V A$ we obtain 
\begin{equation}\label{susikN}
 V_t = V_xA^2 + VA_x A + ({\cal D}_2V) ({\cal D}_1A^2 ) + ({\cal D}_1V)({\cal D}_2A^2)
\end{equation}

This  is our supersymmetric extension of the Novikov equation.

In order to compute the bosonic sector of this equation let us parametrise the superfunctions $V$ and $A$ as

\begin{eqnarray} 
&& V=v_0+\theta_2 \theta_1 v_1, ~~~~~~~ A = u + \theta_2 \theta_1 a_1 , ~~~~~~~ v_0 = u  - 2a_1, \\ \nonumber 
&& v_1=a_1 - \frac{1}{2} u_{xx} ,~~~~ a_1=\frac{1}{2} ( u - \rho) , ~~~~
 v_1 = \frac{1}{2}(u - u_{xx} - \rho)=\frac{1}{2}(m - \rho).
\end{eqnarray}

Due to it the equation (\ref{susikN}) reduces to the two-component generalization of the Novikov equation
\begin{eqnarray}\label{susnowy}
 && \rho_t=\rho_xu^2 + \rho uu_x, \\ \nonumber 
  && m_t = 3u_xum + u^2m_x - \rho(u\rho)_x. 
\end{eqnarray}
 which is different than two-peakon equation (\ref{nu}).

This  equation can be rewritten in the Hamiltonian form as 
\begin{equation}
 \left (\begin{array}{cc} \rho \\ m \end{array} \right )_t = \hat {\cal K} 
\left (\begin{array}{cc} \frac{\delta H}{\delta \rho} \\ \frac{\delta H}{\delta m} \end{array} \right )
\end{equation}
where $\hat {\cal K}$ is a bosonic part of the operator ${\cal K}$ and 
\begin{eqnarray}\nonumber 
&& \hspace{1cm} H = \frac{1}{2}\int ~dx (mu - \rho^2) \\ \nonumber 
&& \hat {\cal K} = 
\left ( \begin{array}{cc}
\rho^{-1}\partial \rho^2 \hat {\cal L}^{-1} \rho^{2}\partial \rho^{-1} ~~~~
 & 3 \rho^{-1}\partial \rho^{2}  \hat {\cal L}^{-1}m^{1/3} \partial m^{2/3}
 \\3m^{2/3} \partial \hat {\cal L}^{-1} \rho^2\partial \rho^{-1}   & 
-\rho \partial \rho + 9m^{2/3}\partial m^{1/3}  \hat {\cal L}^{-1} m^{1/3} \partial m^{2/3}
\end{array} \right )
\end{eqnarray}

where $\hat{\cal L} = \partial^3  - 4\partial_x$. 

Let us notice that $\hat {\cal K}$  operator is a Dirac reduced version of the following Hamiltonian operator 
\begin{eqnarray}\nonumber 
\left ( \begin{array}{ccc}
\partial_{xxx} - 4u\partial - 2u_x & \rho_x - \rho \partial_x & -m_x - 3m\partial
 \\- \rho\partial_x - 2\rho_x  &  0 & 0 \\
-2m_x - 3m\partial_x & 0 & -\rho\partial_x \rho 
\end{array} \right )
\end{eqnarray}
when $u=1$. Now it is easy to verify that this operator satisfy the Jacobi identity and hence $\hat {\cal K}$ satisfy  Jacobi identity as well. It is our decompression idea. 

We have been not able to find the first Hamiltonian structure for this two-component Novikov  equation. On the other side Li and Liu have obtained the first Hamiltonian structure for the Novikov equation using the Bi-Hamiltonian structure of the two-component Novikov equation. In the next section we  explain why this idea  does not work in our case.

\section{Double Extended  Novikov Equation}

\hspace{0.4cm} Our aim is to construct the Bi-Hamiltonian structure for the two-component generalizations of the two-pekon  Novikov equation (\ref{nu}). 
To end this let us notice that Li and Liu \cite{Liu}  defined the second Hamiltonian operator for the two-peakon  generalization of the Novikov equation (\ref{nu}) as
\begin{eqnarray} 
 && \frac{1}{2} \left ( \begin{array}{cc} 3m\partial + 2m_x ~, & 3n\partial + 2n_x \end{array} \right )^{T}
\hat {\cal L}^{-1} \left ( \begin{array}{cc} 3m\partial + m_x ~,  & 3n\partial + n_x \end{array} \right ) 
\\ \nonumber 
&& \hspace{3cm}+ \frac{3}{2} \left ( \begin{array}{cc} m\partial^{-1}m & -m\partial^{-1} n \\
-n\partial^{-1} m & n\partial^{-1}n \end{array} \right )
\end{eqnarray}
This operator is a Dirac reduced version of the following operator
\begin{equation}\label{LiLiuham}
 \frac{1}{2} \left ( \begin{array}{ccc}
-\partial_{xxx} + 4u\partial_x + 2u_x   & 3m^{2/3}\partial_x m^{1/3}   & 3n^{2/3} \partial_x n^{1/3} \\ 
3m^{1/3} \partial_x m^{2/3}  & 3m\partial^{-1}m & -3m\partial^{-1}n    \\ 
3n^{1/3}\partial_x n^{2/3} & -3n\partial^{-1}m   & 3n\partial^{-1} n
\end{array} \right ) 
\end{equation}
when $u=1$ and it satisfies the Jacobi identity. 

We would like to generalize this operator to the higher dimensional case. The supersymmetric approach is less useful for this aim, because we do not know how to find the supersymmetric counterpart of the nonlocal part of  the operator (\ref{LiLiuham}). For this pourpose  we generalize the operator (\ref{LiLiuham}) to  the five dimensional matrix operator

\begin{equation}
  {\cal J} = 
\frac{1}{2} \left ( \begin{array}{ccccc}
-\partial_{xxx} + 4u\partial_x + 2u_x & \rho_{1}^{2}\partial \rho_{1}^{-1}  & 3m_{1}^{2/3}\partial m_1^{1/3}  & 
 \rho_{2}^{2} \partial_x \rho_2^{-1} & 3m_2^{2/3} \partial_x m_2^{1/3} 
 \\ \rho_1^{-1}\partial \rho_1^2  &  0 &  0 & 0 & 0   \\
3m_1^{1/3} \partial_x m_1^{2/3}  &  0 &  {\cal J}_{3,3}  & 0 & {\cal J}_{3,5}  \\ 
\rho_2^{-1}\partial_x\rho_2^2  & 0 & 0  & 0 & 0 \\ 
3m_2^{1/3}\partial_x m_2^{2/3} & 0 & -{\cal J}_{3,5}^{\star}  & 0  & {\cal J}_{5,5}
\end{array} \right )  \nonumber 
\end{equation}
where $u,m_i,\rho_i, i=1,2$  are the function of $t,x$ and 
\begin{eqnarray} 
 {\cal J}_{3,3} &=& \lambda_1 m_1\partial^{-1}m_1 + \lambda_2 m_2\partial^{-1}m_2 + 
\lambda_3 (m_1\partial^{-1}m_2 + m_2\partial^{-1}m_1)  \\ \nonumber 
&& \hspace{1cm}  + k_1\rho_1\partial \rho_1 + 
k_2\rho_2\partial \rho_2 +k_3(\rho_1\partial \rho_2 + \rho_2\partial \rho_1) .\\ \nonumber 
 {\cal J}_{3,5} &=& \lambda_4 m_1\partial^{-1}m_1 + \lambda_5 m_2\partial^{-1}m_2 + 
\lambda_6 m_1\partial^{-1}m_2 + \lambda_7 m_2\partial^{-1}m_1  \\ \nonumber 
&&  + k_4 \rho_1 \rho_2 \partial +k_5 \rho_{1,x} \rho_2 + k_6\rho_1\rho_{2,x} +k_7\rho_1^2 \partial 
+ k_8\rho_2^2\partial \\ \nonumber 
&& \hspace{1cm} +  k_9 \rho_1\rho_{1,x} + k_{10} \rho_2\rho_{2,x}  \\ \nonumber 
{\cal J}_{5,5} &=& \lambda_8 m_1\partial^{-1}m_1 + \lambda_9 m_2\partial^{-1}m_2 + 
\lambda_{10} (m_1\partial^{-1}m_2 + m_2\partial^{-1}m_1)  \\ \nonumber 
&& \hspace{1cm} + k_{11}\rho_1\partial \rho_1 + k_{12}\rho_1\partial \rho_2 +
k_{13}(\rho_1\partial \rho_2 + \rho_2\partial \rho_1) 
.\\ \nonumber 
\end{eqnarray}
and $\lambda_i, k_i$ are an arbitrary constants. 

We would like to consider the following equation of motion 

\begin{eqnarray}\label{4nowy}
&& \left (\begin{array}{cccc} \rho_1 \\ m_1 \\ \rho_2 \\ m_2 \end{array} \right )_t = 
{\cal K} \left ( \begin{array}{cccc} H_{1,\rho_1} \\ H_{1,m_1} \\  H_{1, \rho_2} \\ 
 H_{1,m_2} \end{array} \right ) \\ \nonumber 
H_{1} &=&\frac{1}{2}\int ~dx ~ (m_1u_1 + m_2u_2 + 2\rho_1^2 + 2\rho_2^2) 
\end{eqnarray}
where opertaor ${\cal K}$ is the Dirac reduced version of the operator ${\cal J}$ when $u=1$. 
We assume that  $\lambda_1=\lambda_3=\lambda_4=\lambda_5=\lambda_6=\lambda_9=\lambda_{10}=0$
because then we obtain the local equation of motion.
One can check (see Appendix for details) that the Jacobi identity for the operator ${\cal J}$ holds if 

\begin{eqnarray} 
 {\cal J}_{3,3} &=& s_0 m_2\partial^{-1}m_2  + \rho_1\partial (s_1\rho_1 + s_2\rho_2) + 
\rho_2\partial(s_2\rho_1 + s_3\rho_2)\\ \nonumber 
 {\cal J}_{3,5} &=& -s_0 m_2\partial^{-1}m_1 + s_4(\rho_1\partial \rho_2 - \rho_2\partial \rho_1)  \\ \nonumber 
{\cal J}_{5,5} &=& s_0 m_1\partial^{-1}m_1  + \rho_1\partial (s_1\rho_1 + s_2\rho_2) + 
\rho_2\partial(s_2\rho_1 + s_3\rho_2)
\end{eqnarray}
where now $s_i$ are an arbitrary constants. 

In order to fix the constants $s_i$ we postulate that the system (\ref{4nowy}) is the Bi-Hamiltonian. 
Our result is that for  $s_0=3,s_4=1,~ s_1=s_2=s_3 = 0 $ it is possible to construct the following Bi-Hamiltonian structure

\begin{equation} \label{biha}
 \left ( \begin{array}{cccc} \rho_1 \\ m_1 \\ \rho_2 \\ m_2 \end{array} \right )_t = 
{\cal L} \left ( \begin{array}{cccc}  H_{0,\rho_1} \\
  H_{0, m_1} \\ H_{0, \rho_2} \\ H_{0, m_2} \end{array} \right ) = 
{\cal K} \left ( \begin{array}{cccc} H_{1,\rho_1} \\ H_{1,m_1} \\  H_{1, \rho_2} \\ 
 H_{1,m_2} \end{array} \right ) .
\end{equation}
where 
\begin{eqnarray}
H_{0} &=& -\int ~dx~ m_{1}(u_{2,x}u_1^2 - u_{1,x}u_1u_2) + m_2(u_{2,x}u_1u_2 -u_{1,x}u_2^2)  \\ \nonumber 
&& \hspace{1.5cm} 	+\rho_1\rho_2(u_2u_{2,x} + u_1u_{1,x}) + \rho_2\rho_{1,x}(u_1^2 +u_2^2)  
\end{eqnarray} 

\begin{equation} 
 {\cal L} = \left ( \begin{array}{cccc} 
0 & 0 & 1 & 0 \\
0 & 0 & 0 & 1 - \partial_{xx} \\
-1  & 0 & 0 & 0 \\
0,  & -1 + \partial_{xx}   & 0 & 0\end{array} \right )
\end{equation}

\begin{eqnarray} 
&& {\cal K}= - \frac{1}{2}\left ( \begin{array}{cccc} 
2\rho_{1,x} + \rho_1 \partial \\ 2m_{1,x} + 3m_1\partial \\ 2\rho_{2,x}+\rho_2 \partial \\ 2m_{2,x} + 3m_2 \partial  \end{array} \right )(\partial_{xxx} - 4\partial_x)^{-1}
\left ( \begin{array}{cccc} 
-\rho_{1,x} + \rho_1 \partial \\ m_{1,x} + 3m_1\partial \\ -\rho_{2,x}+\rho_2 \partial \\ m_{2,x} + 3m_2 \partial 
\end{array} \right )^{T}  +\\ \nonumber 
&&  \frac{1}{2}\left ( \begin{array}{cccc} 0 & 0 & 0 & 0  \\
0 & 3 m_2\partial^{-1} m_2 & 0 & - 3 m_2\partial^{-1}m_1 + \rho_2\partial \rho_1 - \rho_1\partial \rho_2 \\
0 & 0 & 0 & 0\\ 
0 & -3 m_1\partial^{-1}m_2 + \rho_1\partial\rho_2 - \rho_2\partial \rho_1 & 0 & 3m_1\partial^{-1}m_1 \end{array} \right )
\end{eqnarray}

The system of equation (\ref{biha}) becomes
\begin{eqnarray}\label{fourNow}
&&   \rho_{1,t} = \rho_{1,x}(u_1^2 + u_2^2) + \rho_1(u_{1,x}u_1 + u_2u_{2,x}) \\ \nonumber 
&&  \rho_{2,t} = \rho_{2,x}(u_1^2 + u_2^2) + \rho_2(u_{1,x}u_1 + u_2u_{2,x}) \\ \nonumber 
&& m_{1,t}= [m_1(u_1^2 + u_2^2)]_x + m_1(u_{1,x} u_1 + u_{2,x}u_2) - 3m_2(u_{2,x}u_1 - u_{1,x}u_2) \\ \nonumber 
&& \hspace{3.5cm} + u_2(\rho_2\rho_{1,x} - \rho_1\rho_{2,x}) \\ \nonumber 
&& m_{2,t}=[m_2(u_1^2 + u_2^2)]_x + m_2(u_{1,x} u_1 + u_{2,x}u_2) + 3m_1(u_{2,x}u_1 - u_{1,x}u_2) \\ \nonumber 
&& \hspace{3.5cm} + u_1(\rho_{1}\rho_{2,x} - \rho_{2} \rho_{1,x})
\end{eqnarray}

It is our double extended  cubic peakon equation. 

We checked using the symbolic computer program that the Hamiltonian  operators ${\cal K}$and ${\cal L}$ are compatible, that is we verivied that 
\begin{equation} 
\int ~ dx ~ f ({\cal K}^{'}[{\cal L}g]  + {\cal L}^{'}[{\cal K}g]) h  + c.p. =0
\end{equation}
where $f,g,h$ are the test function while ${\cal K}^{'}[{\cal L}g]$ denotes the Gateaux derivative along ${\cal L}g$.

It is impossible to reduce double extended  cubic peakon equation (\ref{fourNow})  to the two-component Novikov equation (\ref{susnowy}). Notice that when $\rho_1=\rho_2, m_1=m_2$ we obtain the decoupled system of equation. It is a reason that the idea of Li and Liu does not work in our case.

Let us consider the following   linear transformation  of $m_1,m_2,u_1,u_2,\rho_1,\rho_2$ 
\begin{eqnarray} 
&&  m_1=i(n_1 - n_2)/2, ~~~~ u_1 = i(v_1 - v_2)/2 , ~~~~ \rho_1=ir_1\\ \nonumber 
&&  m_2=(n_1 + n_2)/2, ~~~~ u_2=(v_1 +v_2)/2, ~~~~ \rho_2=r_2
\end{eqnarray}
Under this transformation our equations (\ref{fourNow}) are 
\begin{eqnarray}
&& r_{1,t} = \frac{1}{2} r_1(v_1v_2)_x  + r_{1,x}v_1v_2 \\ \nonumber 
&& n_{1,t}=v_1v_2n_{1,x} + 3v_{1,x}v_2n_1 + v_2(r_2 r_{1,x} - r_1r_{2,x})  \\ \nonumber 
&& r_{2,t} = \frac{1}{2} r_2(v_1v_2)_x  + r_{2,x}v_1v_2 \\ \nonumber 
&& n_{2,t}= v_1v_2n_{2,x} + 3v_{2,x} v_1 n_2  - v_1 (r_2 r_{1,x} - r_1 r_{2,x})
\end{eqnarray}

When $r_1=r_2=0$ then these equation are reduced to the equations considered by Geng and Xue \cite{Geng}.

\section{Conclusion}

 In this paper we constructed the $N=2$ supersymmetric  Hamiltonian structure for  the supersymmetric  Novikov equation.  The bosonic sector gives us the two-component generalization of the cubic peakon equation. Next  we decompressed  second Hamiltonian operator of the two-peakon equation to the five dimensional matrix operator. We checked the Jacobi identity for this operator and reduced this operator to the four dimensional matrix operator. This four dimensional matrix operator was used to the construction of 
the double extended, two-component and two-peakon  Novikov equation.
The first Hamiltonian structure have been defined also and thus  Bi-Hamiltonian structure for this extended system was defined. Moreover these Hamiltonian oprators are compatible. This doubly  extended Novikov equation, up to our knowledge,  is a new Bi-Hamiltonian system. From that reason it is interesting to study it in more details and it is tempting  to check whether this system possess the Lax representation.

\section{Appendix}
We used symbolic computer algebra  for the verification of  the Jacobi identity. 

In  order to proof that the operator $\cal J$   satisfy the Jacobi identity we utilize the standard form of the Jacobi identity \cite{blacha}
\begin{equation} \label{Jacob}
  \int~dx A {\cal J}^{\star}_{\cal J B} C + p.c.=0
\end{equation}
where $A,B,C$ are the test vector  functions as for example  $A=(a_1,a_2,a_3,a_4,a_5 )$ while $\star$ denotes the Gateaux  derivative along the vector $\cal J B$ . 

This formula has three typical  components. 

The first  component contains  terms  in which  the integral operator appear twice,  the second contains terms in which  integral operator appear only once. 
The last third term does not contain the integral operators.

The first component is constructed as 
\begin{equation} 
  \int ~dx~ m_i a_j\partial^{-1} m_k b_s \partial^{-1}m_r c_l + .... 
\end{equation}
This can be be transformed to 
\begin{equation} 
  \int ~dx~m_k b_s (\partial^{-1}m_i a_j)(\partial^{-1}m_r c_l) + .... 
\end{equation}
Introducing  the notation 
\begin{equation} 
 m_i a_j = Z(a,i,j)_x, ~~~ m_i b_j = Z(b,i,j)_x, ~~~ m_ic_j=Z(c,i,j)_x
\end{equation}
the last formula transforms to 
\begin{equation}
 \int ~dx ~ Z(b,k,s)_x Z(a,i,j) Z(c,r,l) + ... 
\end{equation}
Replacing $Z(a,i,j)_x$ by $ \partial Z(a,i,j) - Z(a,i,j) \partial $ 
the first component turn to zero.

The second component is constructed as
\begin{equation}
\int ~dx~ (W_1 + W_2)\partial^{-1} m_ia_j  +  m_ia_j \partial^{-1} (V_1 + V_2) + ... 
\end{equation}
where $W_1$ or $ V_1$ are the  functions constructed out of $\{ \rho_1,\rho_2,\rho_{1,x},\rho_{2,x} b_s,b_{s,x},c_k,c_{k,x}\}$ while  $W_2$ or $V_2$ are constructed out of 
$\{m_i,m_{i,x},c_k,c_{k,x},b_s, b_{s,k}\} $. 
These terms we order as
\begin{eqnarray}
\int ~dx (W_1 + W_2)\partial^{-1} m_ia_j  - (V_1 + V_2) \partial^{-1}  m_ia_j  + ... 
\end{eqnarray}
If we replace  $a_{k,x}$ by $\partial a_k - a_k \partial $  and next  $b_{k,x}$ 
 as $ \partial b_k - b_k\partial $ then the  second component does not contain the integral operators. 

We add just computed  second component to the third component and this sum  vanishes. 
This finish the proof. 

\end{document}